# Coupling-Enhanced Broadband Mid-Infrared Light Absorption in Graphene Plasmonic Nanostructures

Bingchen Deng[†], Qiushi Guo[†], Cheng Li[†], Haozhe Wang[‡], Xi Ling[⊥], Damon B. Farmer[§], Shu-jen Han[§], Jing Kong[‖], and Fengnian Xia[*,†]

[†]*Department of Electrical Engineering, Yale University, New Haven, Connecticut 06511, United States*

[‡]*Department of Mechanical Engineering, Massachusetts Institute of Technology, Cambridge, Massachusetts 02139, United States*

[⊥]*Department of Chemistry, Boston University, Boston, Massachusetts 02551, United States*

[§]*IBM T. J. Watson Research Center, Yorktown Heights, New York 10598, United States*

[‖]*Department of Electrical Engineering and Computer Science, Massachusetts Institute of Technology, Cambridge, Massachusetts 02139, United States*

**ABSTRACT:** Plasmons in graphene nanostructures show great promise for mid-infrared applications ranging from a few to tens of microns. However, mid-infrared plasmonic resonances in graphene nanostructures are usually weak and narrow-banded, limiting their potential in light manipulation and detection. Here we investigate the coupling among graphene plasmonic nanostructures and further show that by engineering the coupling, enhancement of light-graphene interaction strength and broadening of spectral width can be achieved simultaneously. Leveraging the concept of coupling, we demonstrate a hybrid 2-layer graphene nanoribbon array which shows 5 to 7% extinction within the entire 8 to 14 μm (~700 to 1250 $cm^{-1}$) wavelength range, covering one of the important atmosphere "infrared transmission windows". Such coupled hybrid graphene plasmonic nanostructures may find applications in infrared sensing and free-space communications.

Plasmon is the quantum of collective oscillation of free carriers.[1] Plasmons in noble metals such as gold and silver have been explored extensively for decades.[2-5] Strong light confinement and enhanced light-matter interaction have been achieved primarily in visible and near-infrared wavelength range using plasmons in metals.[6-8] Recently, plasmons in atomically thin graphene[9] have also attracted significant attentions due to their strong light confinement, relatively low loss, tunability by an electric field, and its coverage of less-explored wavelength range from terahertz to mid-infrared.[10-23] In mid-infrared, however, the plasmonic resonances in graphene nanostructures are usually weak and narrow-banded, significantly limiting their capabilities in light manipulation and absorption.[18, 24-25] Although previously the broadening of graphene plasmonic resonances have been reported through the utilization of two kinds of nanoribbons with different graphene layer thicknesses in one device,[24] the resonance intensity is reduced because the filling factor (FF, the ratio of graphene area to the whole device area) of ribbons with each thickness decreases. It is therefore highly desirable to realize a design strategy in which the enhancement of light-graphene interaction strength and the broadening of plasmonic spectral width can be achieved simultaneously.

## RESTULS AND DISCUSSION

### Device Fabrication and Measurement Scheme

Two layers of large-area graphene grown by chemical vapor deposition (CVD) were first transferred successively onto 60-nm-thick diamond-like carbon (DLC) on top of a highly resistive silicon substrate.[26] DLC was used due to its non-polar nature such that plasmonic resonance is not affected by the substrate surface polar phonons.[16] We used two layers of graphene here to enhance the overall plasmonic resonance intensity.[21] Nanoribbons were defined using electron-beam lithography in poly (methyl methacrylate) (PMMA) and the following dry etch by oxygen plasma transferred the patterns from PMMA to graphene. (See details in the Methods section)

An infrared microscope coupled to Fourier-transform infrared spectrometer (FTIR) was used to measure the mid-infrared transmission spectra. The incident light was polarized perpendicular to graphene nanoribbons in order to effectively excite the localized plasmons.[16] We measured the transmission $T$ through the nanoribbon array and the reference transmission $T_0$ through the nearby blank region where the graphene was etched away. The extinction spectrum of the nanoribbon array can therefore be expressed as $1-\frac{T}{T_0}$. A schematic of the device and the measuring scheme are shown in Figure 1a. (See details in the Methods section.)

**Plasmonic Coupling among Single-mode Graphene Nanoribbons**

We first study the coupling among single-mode graphene nanoribbons (ribbons with an identical width). Arrays consisting of 100-nm-wide, periodic graphene ribbons with filling factors (FFs) of 20%, 40% and 60% were first investigated. The inset of Figure 1b shows the scanning electron microscopy (SEM) image of 100-nm-wide graphene nanoribbon arrays with a filling factor of 40%. As the filling factor increases the plasmonic coupling among the ribbons becomes more significant. The extinction spectra of the above mentioned three arrays are plotted in Figure 1b. All the extinction curves show clear single resonance peaks in mid-infrared region, which is consistent with the previous reports.[16-17, 21] As shown in Figure 1b, the peaks undergo a red shift as the ribbons get closer. More importantly, the extinction peak reaches nearly 10% for 60% FF nanoribbons, about 5 times as large as that for 20% FF ribbons. This non-linear dependence of the extinction on the FF, together with the red shift, indicates that the significant coupling does exist among these plasmonic nanoribbons.

The same set of experiments was also performed on 160-nm-wide nanoribbon arrays with filling factors of 20%, 40%, and 60%, respectively. The results are summarized in Figure 1c. Here, both the extinction enhancement and red shift in high FF ribbon arrays are less pronounced than those for 100-nm-wide ones. This is because for 160-nm-wide ribbon arrays, the distance between adjacent ribbons is larger, leading to reduced coupling. Here we want to emphasize that, for all the nanoribbons reported in this work, their extinction spectra always show red shift and enhanced extinction intensity per unit graphene area as the coupling gets stronger. This results from the unique properties of graphene plasmonic nanostructures as will be discussed below. This observation is different from that in coupled metallic nanoparticles, in which the coupling can lead to both red or blue shifts in resonance and the intensity can be suppressed or enhanced, depending on the excitation scheme, the periodicity of structures, the size of particles, *etc*.[27-32]

**Modeling of Plasmonic Coupling**

To account for the observed coupling behaviors, we utilize a coupled-dipole theory[27, 30, 32] and model our system as illustrated in Figure 2a, in which the schematic cross section of a graphene nanoribbon array is shown. Each ribbon is labeled by its sequence number such as $i$, while the period (center-to-center ribbon distance) is $d$. The convention of displacement vectors is chosen such that $\vec{r}_{ij}$ refers to the vector starts from the center of ribbon $j$ and ends at the center of ribbon $i$. When incident light is polarized perpendicular to the ribbons, the ribbons will be polarized and the collective oscillation of free carriers in graphene can be excited. Thus for any $i$, the nanoribbon can be modeled as a dipole with an instantaneous dipole moment $\vec{p}_i$. Since all the graphene nanoribbons are equivalent, we have $\vec{p}_i = \vec{p}_j$ for any $i, j$ and for any temporal moment. The dipole moment $\vec{p}_i = \alpha_p \vec{E}_{\text{act}}$, where $\alpha_p$ is the intrinsic polarizability of a single graphene nanoribbon and $\vec{E}_{\text{act}}$ is the actual electric field this ribbon experiences. If all the nanoribbons are decoupled, $\vec{p}_i$ can be further expressed as $\alpha_p \vec{E}_{\text{in}}$, where $\vec{E}_{\text{in}}$ is the incident field.

However, when taking the coupling among ribbons into account, $\vec{E}_{\text{act}} = \vec{E}_{\text{in}}$ no longer holds. Instead, we have[27, 30, 32]

$$\vec{p}_i = \alpha_p \vec{E}_{\text{act}} = \alpha_p (\vec{E}_{\text{in}} + \sum_{j \neq i} \tilde{c}_{ij} \vec{p}_j), \quad (1)$$

where $\tilde{c}_{ij} \vec{p}_j$ represents the retarded dipole field produced by $\vec{p}_j$ (ribbon $j$) at the center of ribbon $i$. The full expression of $\tilde{c}_{ij} \vec{p}_j$ can be found elsewhere.[33] For simplicity, we use a “nearest-neighbor approximation” in which only two adjacent ribbons interact with each other. Thus Eq. (1) can be simplified as

$$\vec{p}_i = \alpha_p (\vec{E}_{\text{in}} + \tilde{c}_{i,i-1}\vec{p}_{i-1} + \tilde{c}_{i,i+1}\vec{p}_{i+1}), \tag{2}$$

where $\tilde{c}_{i,i-1}\vec{p}_{i-1}$ and $\tilde{c}_{i,i+1}\vec{p}_{i+1}$ originate from the contribution of the ribbon on the left and right of ribbon $i$, respectively. Considering our practical configuration (see Figure 1a) and utilizing $\vec{p}_{i-1} = \vec{p}_{i+1} = \vec{p}_i$, Eq. (2) can be further calculated as

$$\vec{p}_i = \alpha_p \left[\vec{E}_{\text{in}} + \frac{1}{\pi\varepsilon_0 d^3} e^{\text{i}kd}(1-\text{i}kd)\,\vec{p}_i\right], \tag{3}$$

where $\varepsilon_0$ is the vacuum permittivity, $e$ is the base of the natural logarithm, $\text{i}$ the imaginary unit and $k$ is the modulus of the wave vector of the incident light. From (3) we can explicitly obtain $\vec{p}_i$ as

$$\vec{p}_i = \frac{1}{\frac{1}{\alpha_p} - \frac{1}{\pi\varepsilon_0 d^3} e^{\text{i}kd}(1-\text{i}kd)} \vec{E}_{\text{in}} \stackrel{\Delta}{=} \alpha_{\text{eff}} \vec{E}_{\text{in}}, \tag{4}$$

where the effective polarizability $\alpha_{\text{eff}}$ is defined.

In decoupled case, using Lorentz model the polarizability (as the function of angular frequency $\omega$) can be expressed as[34]

$$\alpha_p = \frac{Nq^2}{m_0} \frac{1}{\omega_0^2 - \omega^2 - \text{i}\gamma\omega} = \frac{A}{\omega_0^2 - \omega^2 - \text{i}\gamma\omega}, \tag{5}$$

where $N$ is the number of carriers in a single nanoribbon, $q$ is the elementary charge, $m_0$ is the effective mass of carriers, $\omega_0$ is the intrinsic oscillation frequency, $\gamma$ is the damping rate, and $A = \frac{Nq^2}{m_0}$. The resonance condition $\Re(\alpha_p) = 0$ yields $\omega_{\text{ind}} = \omega_0$, where $\omega_{\text{ind}}$ represents the resonance frequency of an individual nanoribbon. Substituting $\omega_{\text{ind}} = \omega_0$ into (5), we obtain the ribbon polarizability at plasmonic resonance $|\alpha_p|_{\text{ind}} = \frac{A}{\gamma\omega_0}$.

The effective polarizability $\alpha_{\text{eff}}$ as shown in (4) is different from $\alpha_p$ when considering the coupling among nanoribbons in an array. For the graphene plasmonic system studied in this work, the wave vector $k$ is around $\frac{2\pi}{10}\mu\text{m}^{-1}$ while the period of the structure $d$ is of the magnitude of several hundreds of nanometers, therefore $kd \ll 1$ holds. Utilizing this approximation, $\alpha_{\text{eff}}$ is calculated to be

$$\alpha_{\text{eff}} = \frac{1}{\frac{\omega_0^2 - \omega^2}{A} - \frac{1}{\pi\varepsilon_0 d^3} - \text{i}\left(\frac{\gamma\omega}{A} + \frac{k^3}{2\pi\varepsilon_0}\right)}. \tag{6}$$

Again, by applying the resonance condition $\Re(\alpha_{\text{eff}}) = 0$ and using $kd \ll 1$, the resonance frequency of an array

$$\omega_{\text{arr}} = \omega_0 - \frac{A}{2\pi\varepsilon_0\omega_0 d^3} \tag{7}$$

is obtained. Substituting (7) into (6), we obtain the modulus of the effective polarizability of one ribbon in a coupled array:

$$\frac{1}{|\alpha_{\text{eff}}|_{\text{arr}}} = \frac{\gamma\omega_0}{A} - \frac{\gamma}{2\pi\varepsilon_0\omega_0 d^3} + \frac{\omega_0^3}{2\pi\varepsilon_0 c^3} - \frac{3A\omega_0}{(2\pi\varepsilon_0)^2 c^3 d^3}, \tag{8}$$

where $c$ is the speed of light. The third term in (8) is much smaller than the second term considering $\gamma \sim 100\,\text{cm}^{-1}$ (ref. 16), thus can be neglected. Eq. (8) is therefore simplified as:

$$\frac{1}{|\alpha_{\text{eff}}|_{\text{arr}}} = \frac{\gamma\omega_0}{A} - \frac{\gamma}{2\pi\varepsilon_0\omega_0 d^3} - \frac{3A\omega_0}{(2\pi\varepsilon_0)^2 c^3 d^3} \tag{9}$$

It is clear from (7) and (9) that as the nanoribbons get closer, the resonance frequency of an array will be lower while the effective polarizability will increase, which explains the phenomena of red shift and enhanced extinction per unit graphene area. More importantly, here we obtain analytic solutions which describe the effects of coupling in graphene nanoribbon arrays.

The analysis presented above also reveals the reasons why the coupling behaviors in graphene plasmonic nanostructures are different from those in metals. In metal nanostructures, the plasmonic resonance frequencies are usually in visible and near-infrared spectral range.[13, 18] Then $kd$ appears in (4) is of the magnitude of unity. In this case, the term $\frac{1}{\pi\varepsilon_0 d^3} e^{\mathrm{i}kd}(1-\mathrm{i}kd)$ oscillates with the period $d$. Besides, the plasmonic resonance in metal is significantly stronger than that in graphene, thus the "nearest-neighbor approximation" can be invalid.[35] Therefore, both red and blue shifts of the plasmonic resonance frequency and both enhancement and suppression of the resonance intensity were previously observed in coupled metal plasmonic nanostructures.[27-32]

Furthermore, we confirm the validity of Eqs. (7) and (9) using experimental results. From Figures 1b and 1c we extracted the plasmonic resonance frequencies ($\omega_{\mathrm{arr}}$) of nanoribbon arrays with different periods. We denoted these data points in $\omega_{\mathrm{arr}} - d^{-3}$ plot as shown in Figure 2b for both 100-nm and 160-nm results. As predicted by Eq. (7), linear relations are clearly obtained. Each straight line, with two fitting parameters, is applied to fit the experimental data. We also used the normalized extinction (extinction intensity per unit graphene area) to evaluate the quantity $|\alpha_{\mathrm{eff}}|_{\mathrm{arr}}$ in Eq. (9), since the extinction is proportional to the effective polarizability. For example, the normalized extinctions (arbitrary unit) for 20%, 40%, and 60% FF 100-nm-wide nanoribbon arrays are 2, 4.4/2, and 9.5/3, respectively. In $\frac{1}{|\alpha_{\mathrm{eff}}|_{\mathrm{arr}}} - d^{-3}$ plot, we again observed linear relations as shown in Figure 2c. In metals nanoparticles, the red shift is usually described by a simple exponential function of inter-particle distance, due to the complex form of the analytic expression.[36-39] Here in graphene nanoribbons, we reached two simple equations that explicitly describe the coupling effects which scale with $d^{-3}$.

**Plasmonic Coupling among Two-mode Graphene Nanoribbons**

If a graphene plasmonic device consists of nanoribbons with different widths, its operational bandwidth can be broadened. Here we further explore the plasmonics coupling effects among two-mode nanoribbons (ribbon arrays with two different widths) for the enhancement of operational bandwidth and the light-graphene interaction strength simultaneously. In this set of experiments, we first fabricated 160 and 100-nm-wide single-mode graphene nanoribbon arrays with the same period of 640 nm, corresponding to filling factors of 25% and 15.6%, respectively. We further compared their extinction spectra with hybrid graphene nanoribbon arrays, in which we combined the 100-nm-wide and 160-nm-wide ribbon arrays in one device while keeping the same period of 640 nm. Naturally, the hybrid arrays have a filling factor of 40.6% (25%+15.6%). We varied the distance between the 160-nm-wide and the 100-nm-wide ribbons and measured the extinction spectra at different ribbon spacing (see insets in Figure 3a-c for the arrangement). Since there are only one 160-nm-wide ribbon and one 100-nm-wide ribbon within a period of 640 nm, the coupling among the ribbons in different periods is negligibly small.

The extinction spectrum of a hybrid array is represented by the blue curve in Figure 3a, in which the center-to-center distance $G$ between the 160-nm-wide and the 100-nm-wide ribbons is 190 nm, corresponding to 60-nm gap between them. We also plot the extinction spectra of the single-mode 160 and 100-nm-wide graphene nanoribbon arrays with the same period of 640 nm in black and red curves, respectively. It is rather clear that the spectrum of the hybrid array shows much stronger extinction than the sum of two single-mode nanoribbon arrays. Moreover, the red shifts of the two resonance peaks are also observed. These findings are in fact consistent with those observed in the coupled single-mode nanoribbons discussed in the previous section. Since the condition $kG << 1$ ( $k$ is the wave vector and $G$ is the ribbon center-to-center distance) still holds, the retarded dipole field produced by the adjacent ribbon is in phase with the dipole moment of the examined ribbon. As a result, the phenomena observed here in coupled two-mode nanoribbons are qualitatively same with those in coupled single-mode ones. When the center-to-center distance $G$ is 230 nm, the enhancement effect is still evident (see Figure 3b). When $G$ equals or exceeds 280 nm (see Figure 3c for $G$ of 280 nm), the extinction spectrum of the hybrid structure is close to the simple addition of the two extinction spectra measured from two single-mode nanoribbon arrays.

The observed enhancement of the light-graphene interaction is summarized in Figure 3d, in which we plot the integrated overall extinction versus the ribbon spacing $G$. Here, the integrated extinction is obtained by integrating the extinction curve over a spectral range from 600 to 2300 $cm^{-1}$. The integrated extinctions of the hybrid structures are denoted by the pink dots in the figure, and those of the two single-mode nanoribbon structures consisting of uniform widths (160 nm and 100 nm) are represented by the blue and red horizontal dashed lines, respectively. The sum of the integrated extinctions yielded by the two single-mode arrays is also shown by the green dashed line. The integrated extinction increases as the coupling is enhanced, although the overall filling factor (FF) remains at 40.6%. An array with more strongly coupled graphene plasmonic nanostructures hence shows more efficient light absorption per unit graphene area.

**Broadband Mid-infrared Light Extinction Covering the "Infrared Transmission Window".**

Finally we show that coupled plasmonic nanostructures consisting of different widths can rather uniformly cover one "infrared transmission window" spanning from 8 to 14 μm (~700 to 1250 $cm^{-1}$). In the section above, we show that the coupled nanostructures consisting of 160- and 100-nm-wide ribbons exhibit strong light-graphene interaction from ~700 and 1200 $cm^{-1}$. Here in order to slightly extend the spectral coverage to 1250 $cm^{-1}$, we need to reduce the width of 100-nm-wide ribbons slightly. Furthermore, in the extinction spectrum of strongly coupled nanostructures consisting of 160- and 100-nm-wide ribbons (Figure 3a), the second extinction peak (at around 1100 $cm^{-1}$) is much weaker than the first peak. Two factors contribute to this observation. First, the filling factor (FF) of 100-nm-wide ribbons is smaller than that of 160-nm-wide. Second, narrow ribbons have larger damping rate due to more pronounced edge scattering,[16] leading to smaller intrinsic polarizability. As a result, in order to achieve more uniform extinction across the desirable spectral range, the filling factor of narrow ribbons should be greater than that of the wide ribbons. Finally, the spacing among the ribbons should be small enough to facilitate the coupling.

Taking the factors discussed above into account, we fabricate the coupled nanoribbon structures shown in Figure 4a for the uniform coverage of the spectral range between 8 to 14 μm. This structure has a period of 640 nm, in which there are one 160-nm-wide ribbon and three 80-nm-wide ribbons. Moreover, the gap between the adjacent ribbons is kept at 60 nm. This complex hybrid structure yields an extinction spectrum as shown by the blue curve in Figure 4b. It exhibits 5 to 7% extinction from ~700 to 1250 $cm^{-1}$, covering the important atmosphere "infrared transmission window". To illustrate the coupling enhanced light-graphene interaction, the extinction spectra of the two single-mode ribbon arrays that constitute the hybrid structure are also shown by the black and red curves in Figure 4b, respectively. It is obvious that the hybrid structure exhibits much stronger absorption within the entire wavenumber range compared to the simple addition of extinction spectra of two single-mode nanoribbon arrays.

## CONCLUSIONS

We reveal the unique coupling behaviors among mid-infrared localized plasmons in graphene nanostructures, finding that in single-mode periodic nanoribbon arrays both the resonance red shift and the enhancement of the effective polarizability scale with $d^{-3}$, where $d$ is the center-to-center spacing. More importantly, we find that such coupling can effectively enhance the light-graphene interaction and tailor the light-graphene interaction bandwidth. Finally leveraging the coupling among different plasmonic modes, we realize a hybrid graphene nanostructure which strongly interacts with infrared light within one of the "infrared transmission windows" (8 to 14 μm), and such hybrid graphene nanostructures may have applications in sensing and free-space communications.

## METHODS

**Device Fabrication.** Two layers of large-area CVD graphene on copper foil were transferred successively onto 60 nm DLC on highly resistive silicon substrates using the standard wet transfer method.[26] The graphene sheet transferred by this method is hole-doped with a Fermi level of around -0.3 eV,[16] which corresponds to a carrier density of $6.6 \times 10^{12}$ $cm^{-2}$ (ref. 40). The poly (methyl methacrylate) (PMMA) resist layer was spun on top as the dry etch mask. The size of graphene nanoribbon array was designed to be 100 μm × 100 μm, larger than the incident spot size. A Vistec 100-kV electron-beam lithography system was used to define the patterns. Dry etch was performed using oxygen plasma in an Oxford Plasmalab-100 system. The PMMA mask layer was removed in acetone before the transmission measurements.

**Infrared Spectroscopy Measurements.** The mid-infrared spectra were recorded by a Bruker Vertex 70 FTIR spectrometer integrated with a Hyperion 2000 infrared microscope. The polarization of the incident light was adjusted to be perpendicular to the as-fabricated nanoribbons by a holographic wire grid polarizer. During the measurements, the size of the measuring window was kept at 80 μm × 80 μm, smaller than that of an array. The transmissions $T$ and $T_0$ were measured through the graphene nanoribbon array area and the nearby reference area without graphene, respectively. The extinction spectrum $1-\frac{T}{T_0}$ was then obtained.

## AUTHOR INFORMATION

**Corresponding Author**

*E-mail: fengnian.xia@yale.edu.

**Notes**

The authors declare no competing financial interest.

## ACKNOWLEGEMENTS

We thank the Office of Naval Research (N000141410565) and the National Science Foundation (ECCS 1552461) for the support of this work.

**FIGURE CAPTIONS**

**Figure 1. Characterization of single-mode graphene nanoribbon arrays with different filling factors (FF). (a)** Mid-infrared transmission measurement scheme for graphene nanoribbons. The excitation light is broadband and polarized perpendicular to the ribbons. Localized carrier oscillation is denoted as charges accumulating at two ends of ribbons at a certain moment. **(b)** Extinction spectra of 100-nm-wide nanoribbon arrays with filling factors of 20%, 40% and 60%, respectively. Inset: SEM image of 2-layer 100-nm-wide graphene nanoribbon array with 40% filling factor. **(c)** Extinction spectra of 160-nm-wide nanoribbon arrays with filling factors of 20%, 40% and 60%, respectively. Inset: SEM image of 2-layer 160-nm-wide graphene nanoribbon array with 40% filling factor.

**Figure 2. Modeling of plasmonic coupling in single-mode graphene nanoribbon arrays. (a)** Schematic model of the graphene nanoribbon array. Variables used in the main text are indicated. Plasmon oscillation in each ribbon is modeled as a dipole. **(b)** Plasmon resonance frequency as the functions of $d^{-3}$, where $d$ is the period of the ribbon array. Linear relations are shown. Dashed lines are fited based on equation (7). **(c)** Reciprocal of normalized extinction (a.u.) as the functions of $d^{-3}$. Dashed lines are fitted based on equation (9).

**Figure 3. Characterization of coupled two-mode graphene nanoribbon arrays with different coupling stengths but the identical filling factor (FF). (a-c)** Extinction spectra of single-mode 160-nm-wide graphene nanoribbon array (black), single-mode 100-nm one (red), and hybrid ones that consist of the two single-mode arrays (blue). Insets: SEM images of the respective hybrid structures. From (a) to (c), the distance $G$ between the 160-nm-wide ribbon and the 100-nm-wide one in the hybrid array increases

from 190 nm to 280 nm, while the filling factor (FF) remains unchanged. **(d)** The integrated extinction as the function of $G$. For comparison, integrated extinctions of the single-mode 160-nm-wide and 100-nm-wide ribbon arrays and their summation are indicated as the horizontal dashed blue, red, and green lines, respectively.

**Figure 4. Demonstration of a hybrid graphene nanostructure showing enhanced broadband mid-infrared light absorption. (a)** SEM image of the hybrid ribbon array. **(b)** Extinction spectrum of the hybrid structure (blue) together with that of the two single-mode structures that constitute the hybrid (black and red). The atmosphere "infrared window" is indicated by the shadowed region.

**Figure 1. Characterization of single-mode graphene nanoribbon arrays with different filling factors (FF).**

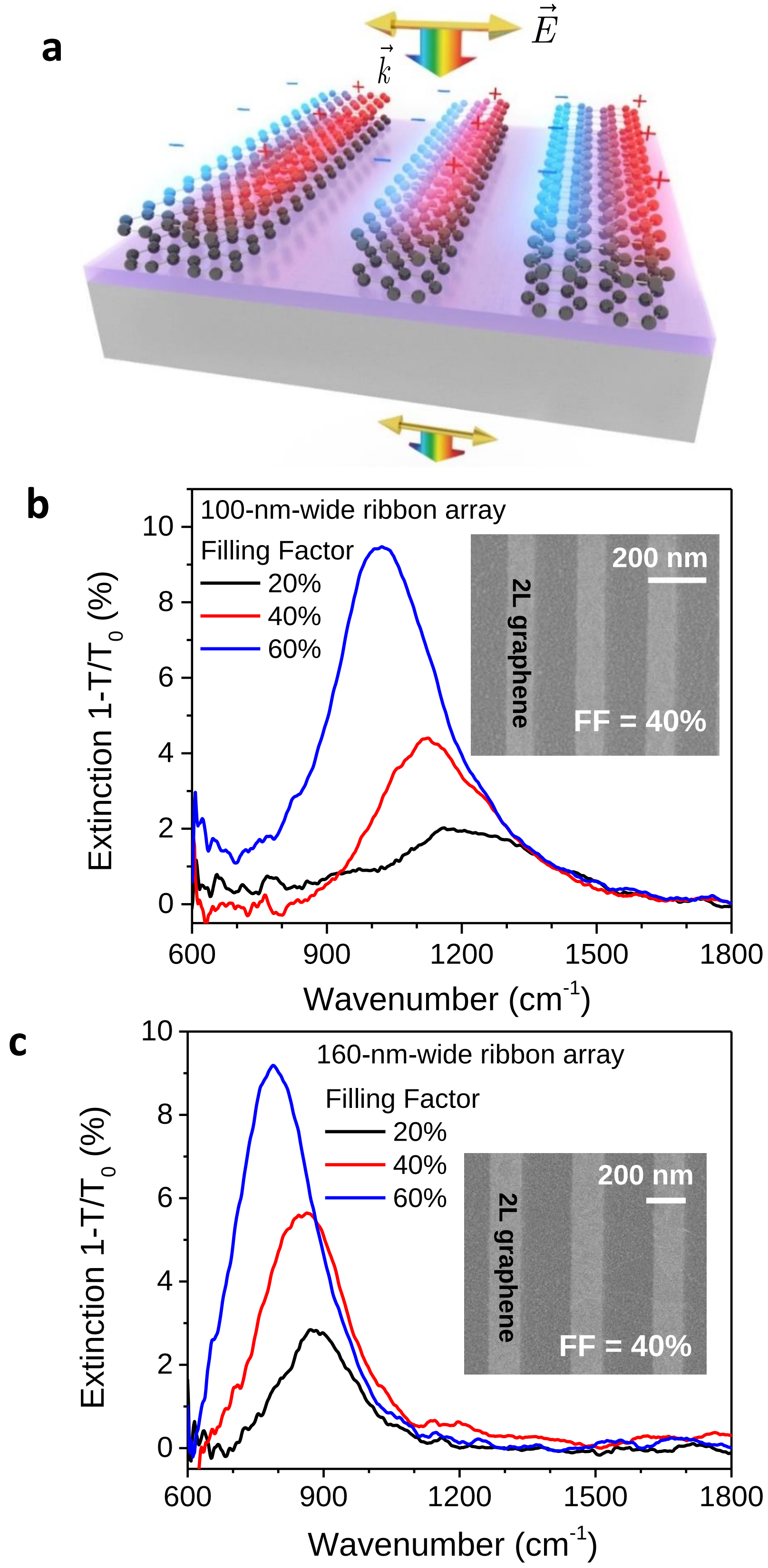

**Figure 2 Modeling of plasmonic coupling in single-mode graphene nanoribbon arrays**

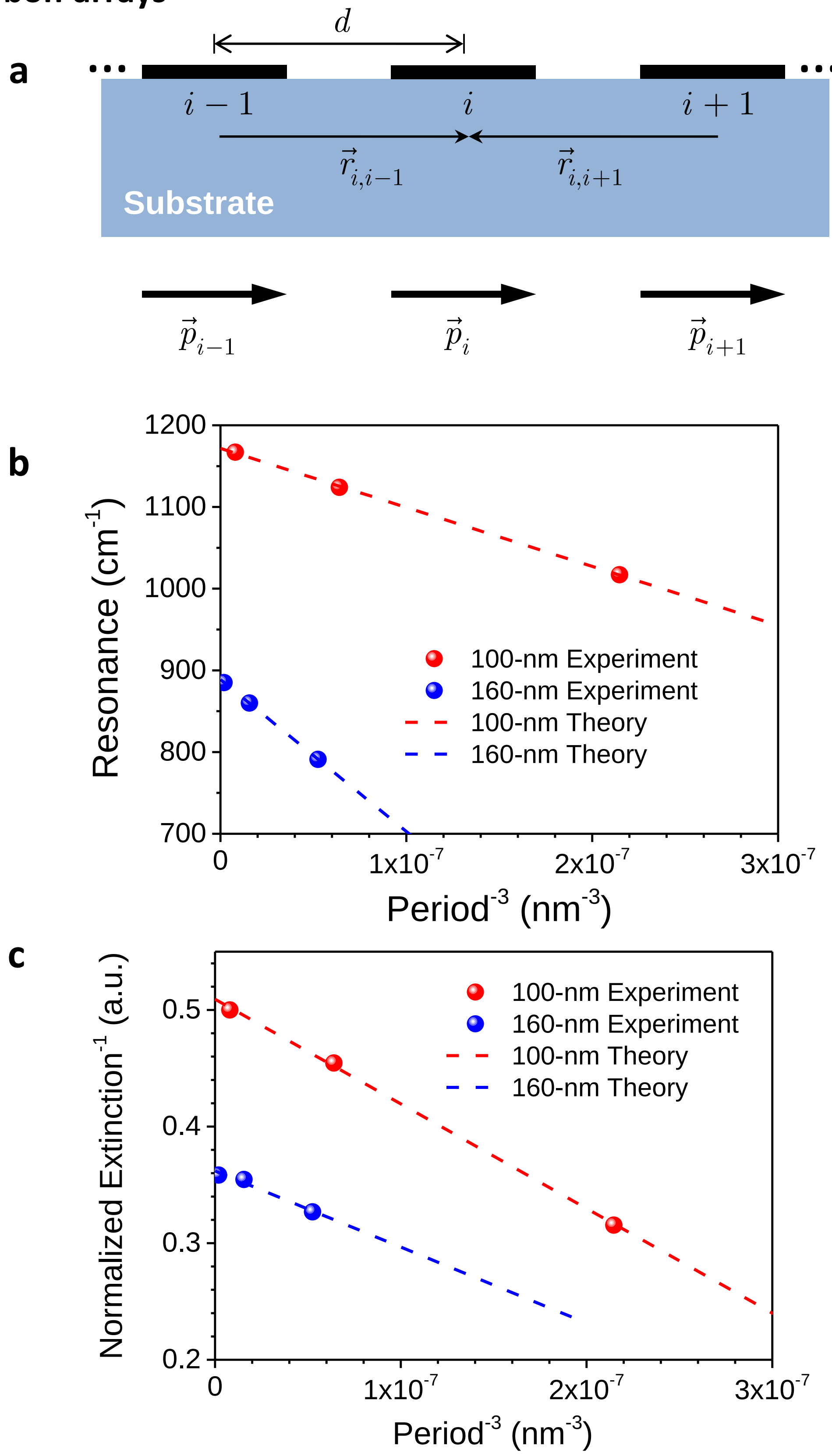

**Figure 3. Characterization of coupled two-mode graphene nanoribbon arrays with different coupling strengths but the identical filling factor (FF)**

a

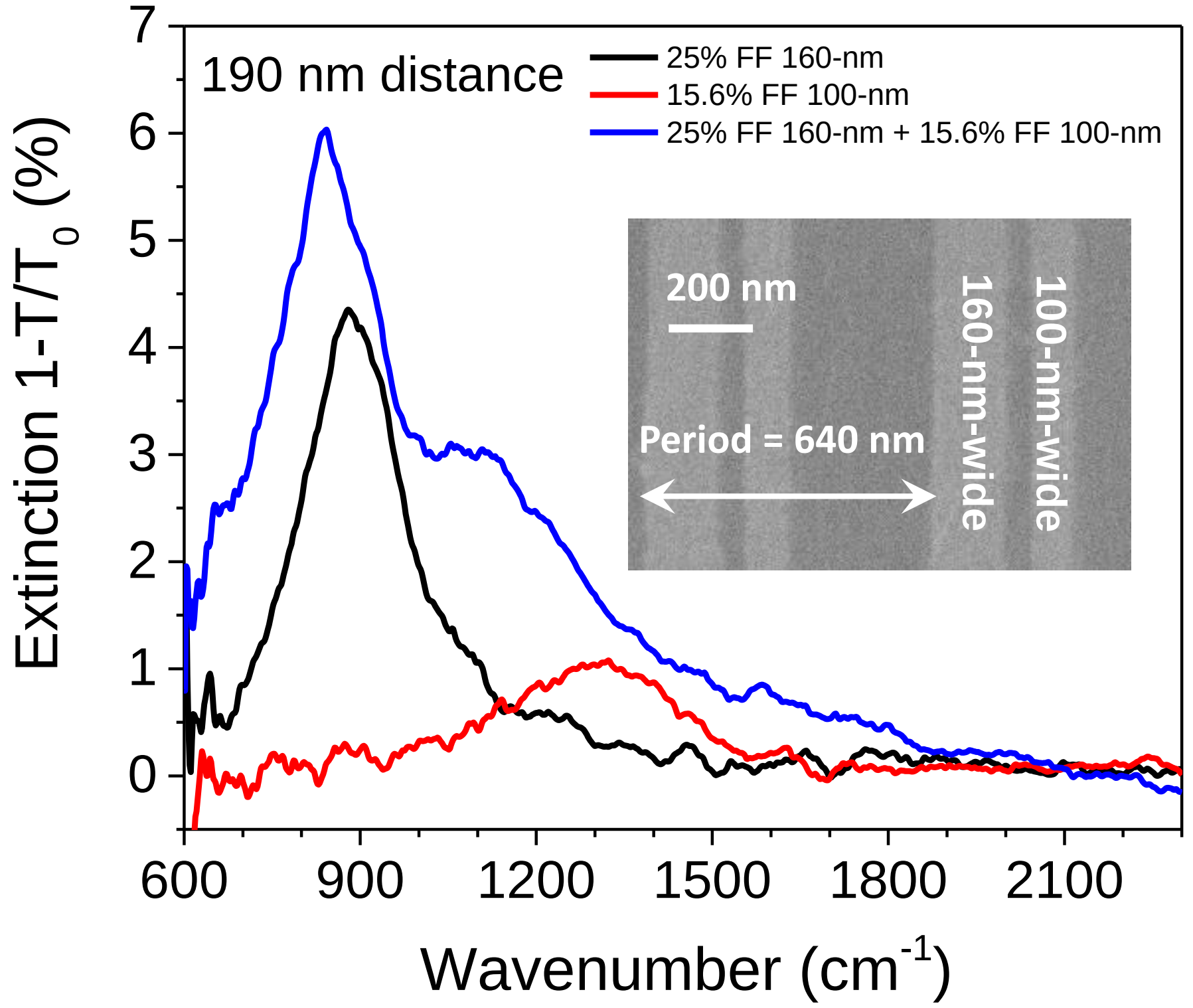

b

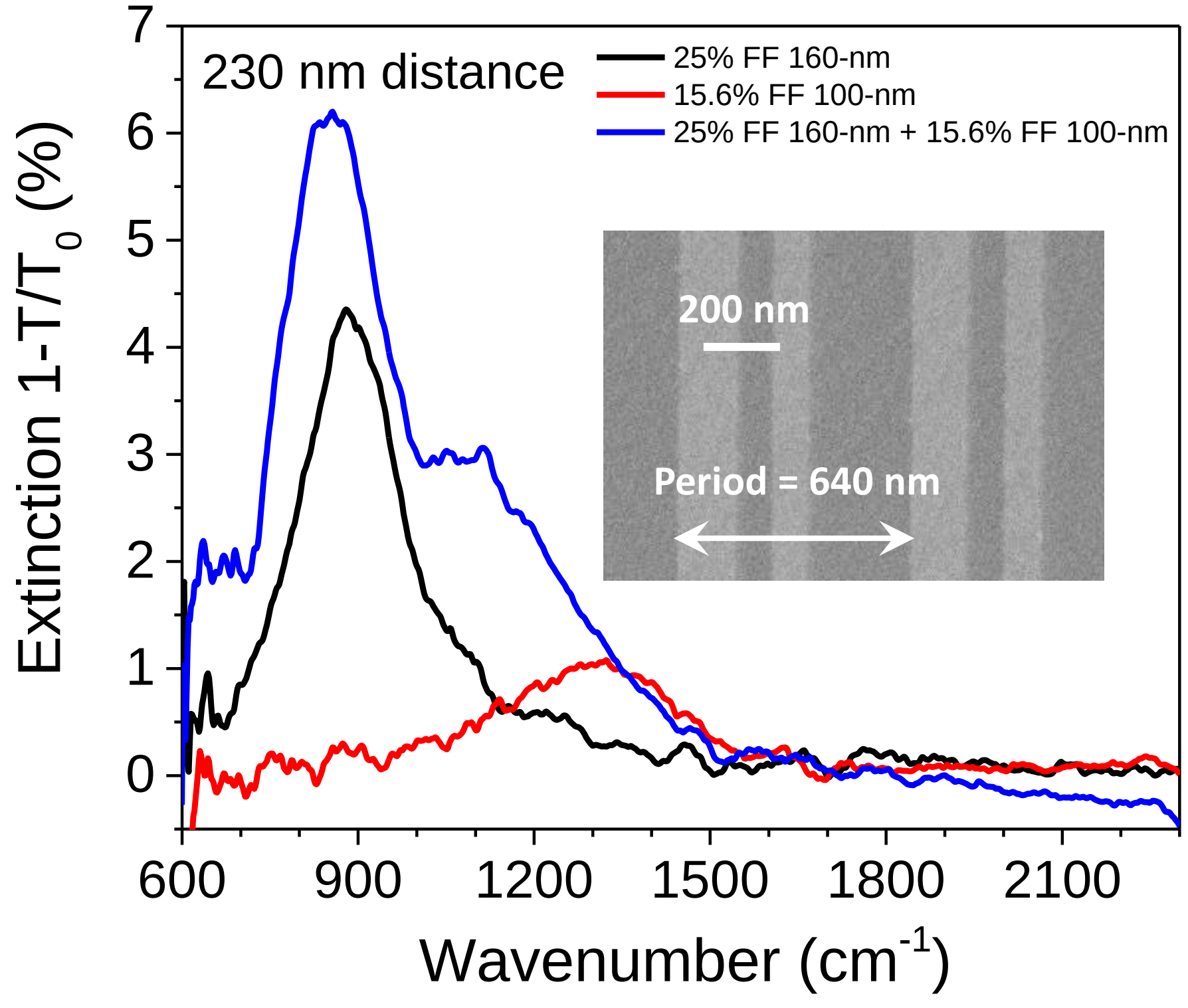

c

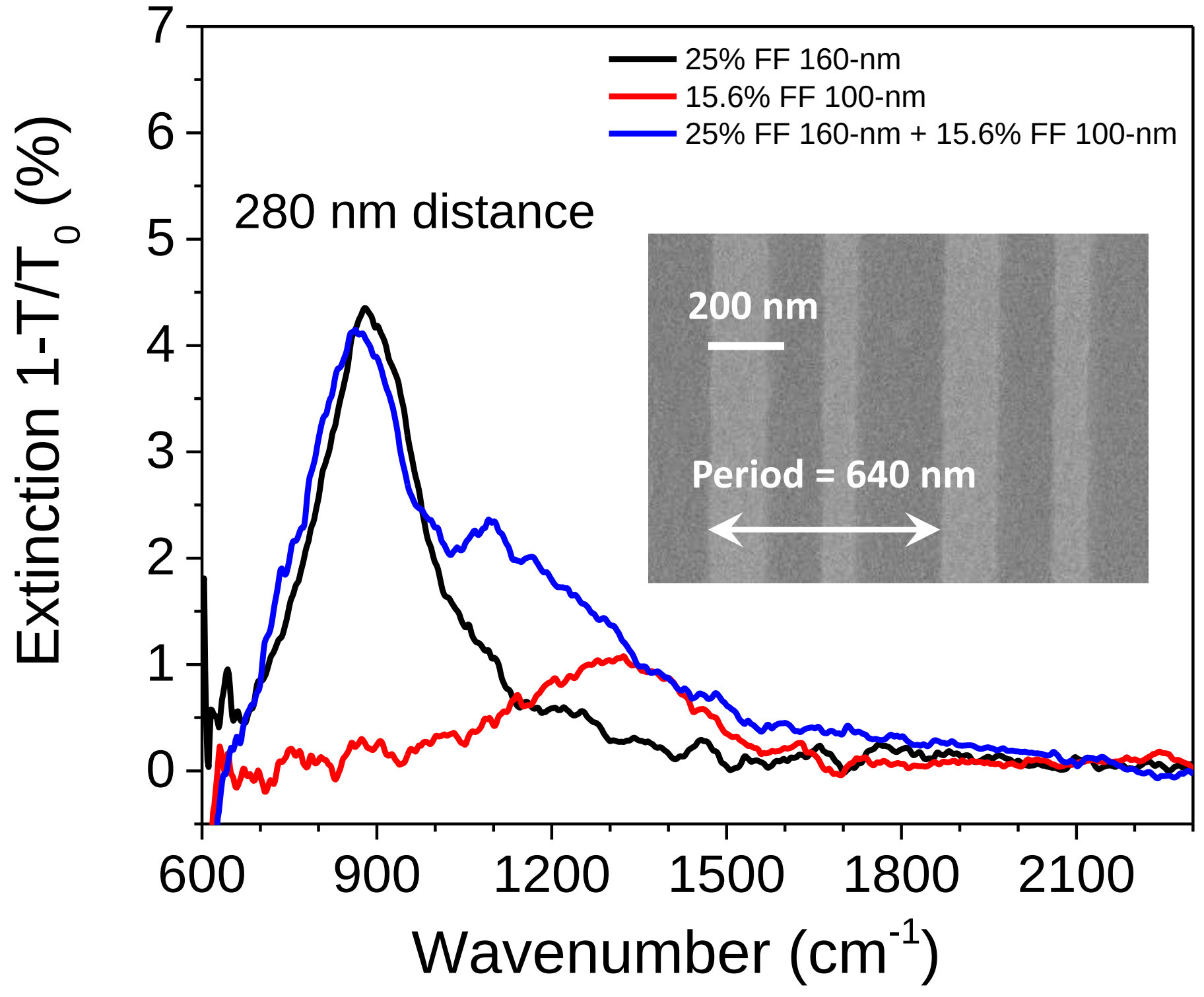

25% FF 160-nm
15.6% FF 100-nm
25% FF 160-nm + 15.6% FF 100-nm
280 nm distance
200 nm
Period = 640 nm
Extinction 1-T/T0 (%)
Wavenumber (cm-1)
0
1
2
3
4
5
6
7
600
900
1200
1500
1800
2100

d

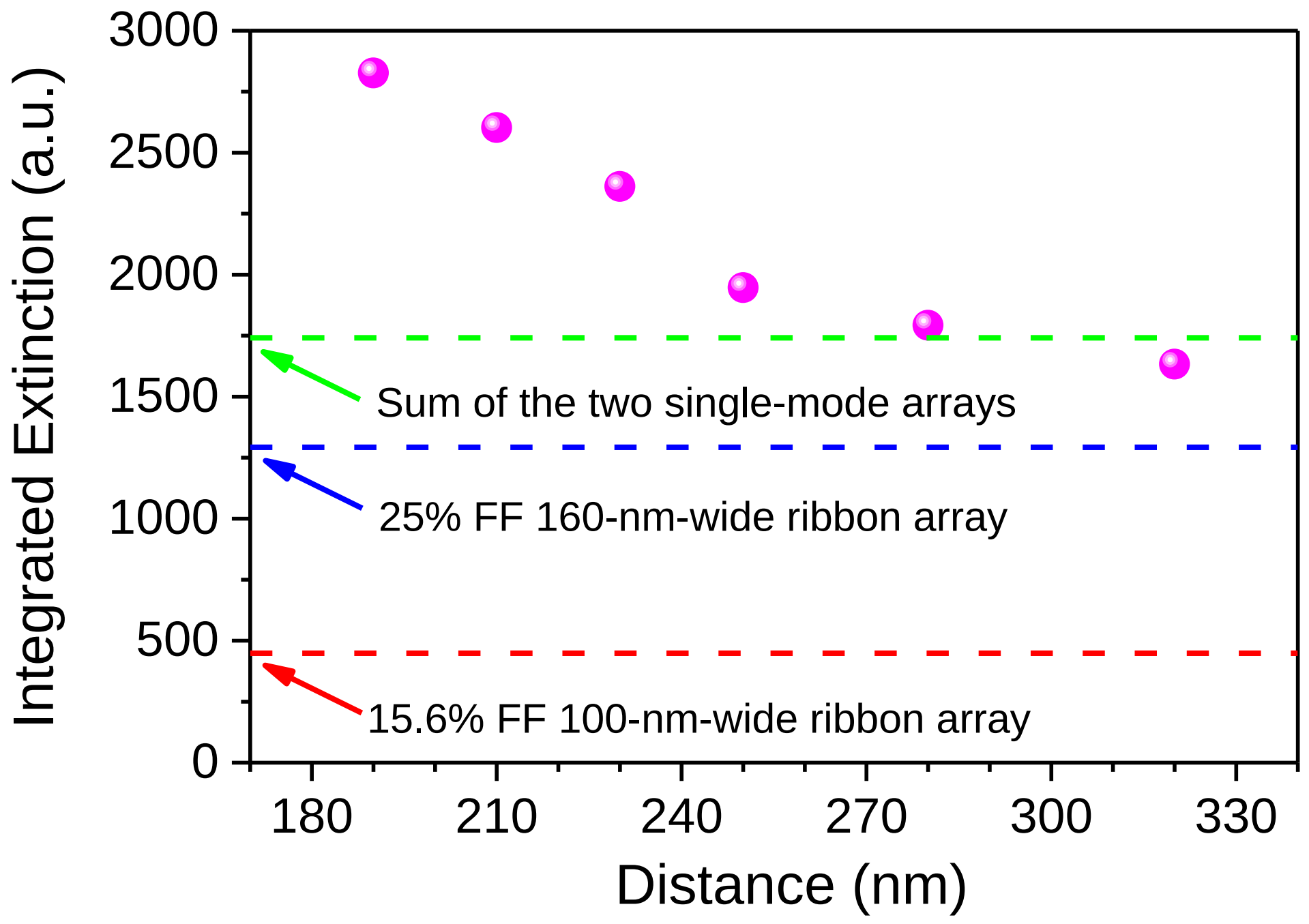

Sum of the two single-mode arrays
25% FF 160-nm-wide ribbon array
15.6% FF 100-nm-wide ribbon array
Integrated Extinction (a.u.)
Distance (nm)
0
500
1000
1500
2000
2500
3000
180
210
240
270
300
330

**Figure 4. Demonstration of a hybrid graphene nanostructure showing enhanced broadband mid-infrared light extinction**

a

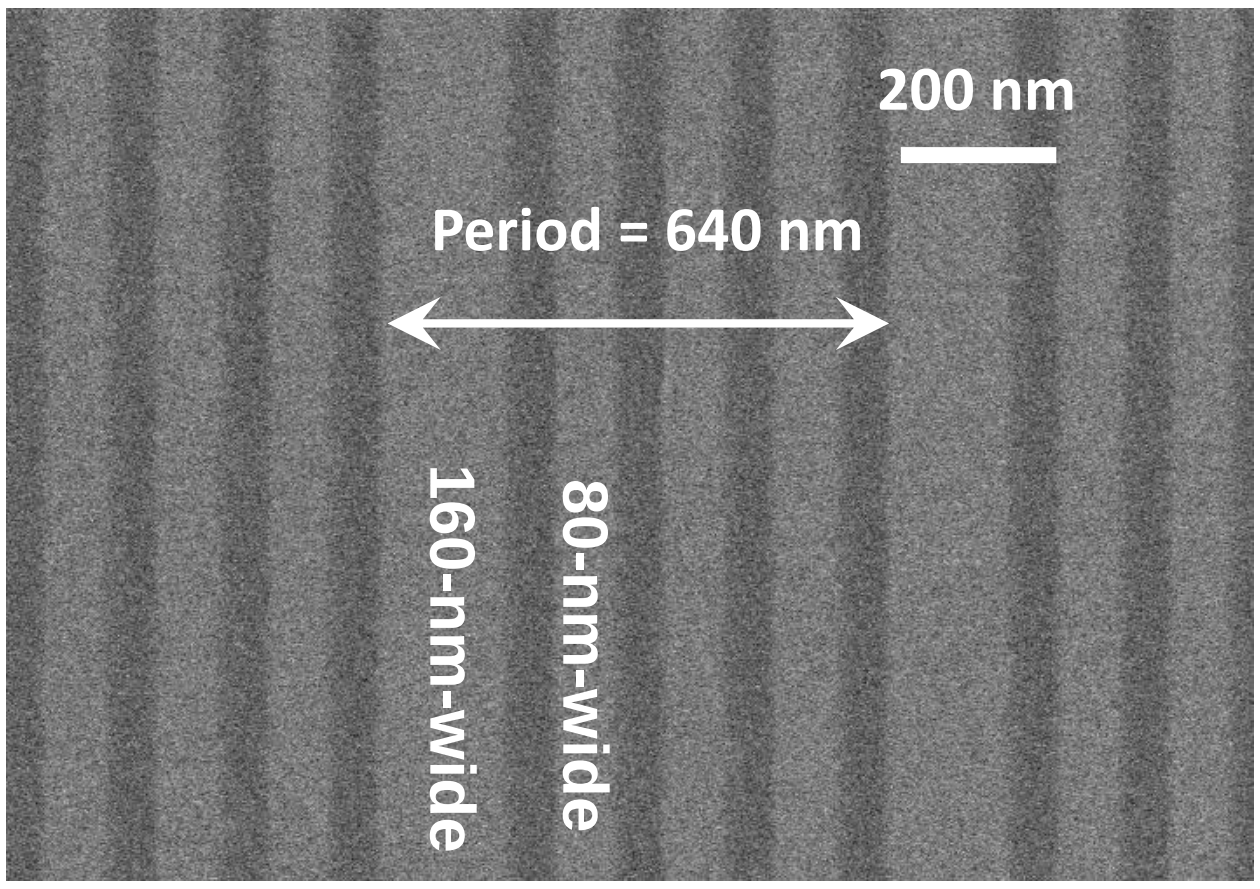

b

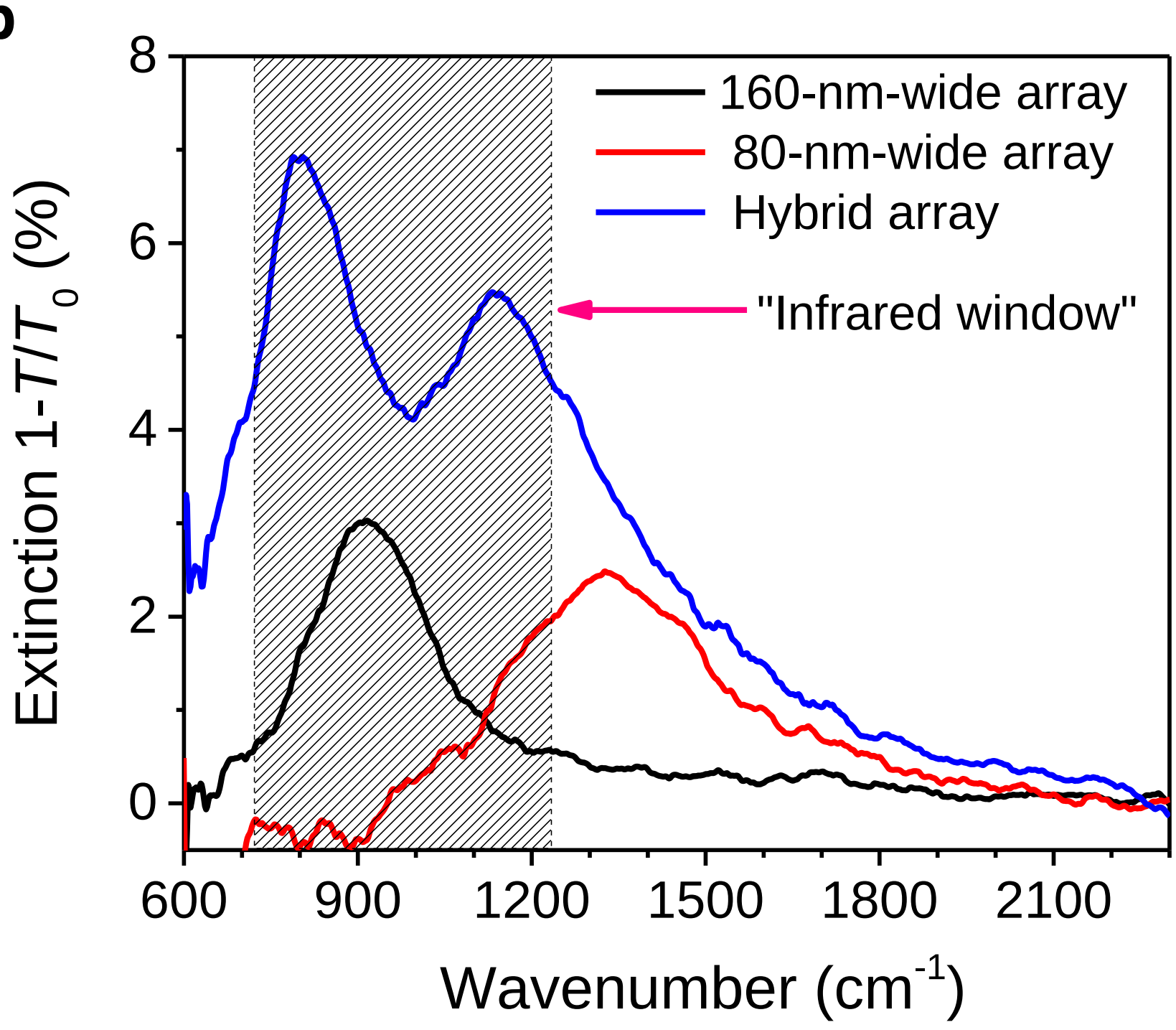